\documentclass{article}
\usepackage{fullpage}
\usepackage{amsfonts,amssymb}
\usepackage{amsthm}
\usepackage{tikz}
\usetikzlibrary{calc}
\usetikzlibrary{arrows}
\usepackage{tikz-3dplot}
\usepackage{tikz}
\usepackage{color} 
\usepackage[fleqn]{amsmath}
\usepackage{amsthm}
\usepackage{amssymb}
\usepackage{amsfonts}
\usepackage{bbm}
\usepackage{epsfig}
\usepackage{times}
\usepackage{hyperref}
\usepackage{bm}
\usepackage{float} 
\usepackage{appendix} 
\usepackage{lscape}
\usepackage{cite}
\usepackage{mathrsfs}
\usepackage{appendix}
\usepackage{setspace}
\usepackage{float}
\usepackage{mathtools}
\usepackage{mdframed}
\usepackage{authblk}
\theoremstyle{definition}

\begin{document}

	\title{How are Entanglement Entropies Related to Entropy Bounds?}

\author{Emily Adlam  \thanks{Philosophy Department and Institute for Quantum Studies, Chapman University, Orange, CA92866, USA \texttt{eadlam90@gmail.com} }}
\date{\today}

	\date{\today} 
	
\maketitle

 \section{Introduction \label{EE}} 
 
 As physically embodied observers, we are subject to various kinds of epistemic limitations, which inevitably have an effect on the way in which we formulate our physical theories. One particularly important kind of epistemic limitation results from the locality of interactions, which means that, in a four-dimensional spacetime,  surfaces of two spatial dimensions have a special role  as interfaces through which observers access physical systems and regions of space. In particular, this means that if there are bounds on the amount of information that can be stored on a surface, there are also bounds on the rate at which we can extract information through those surfaces about the systems we are trying to study, and it is likely that such bounds will have consequences for the physics that we arrive at on the basis of our observations. 
 
 In ref  \cite{PartI}  it was argued that this feature of our  situation as physically embodied observers leads to a possible epistemic interpretation of the covariant entropy bound:  rather than counting the true number of ontological degrees of freedom inside a region of spacetime, the bound could  codify an epistemic limitation pertaining to the amount of information we can extract from a region of spacetime. This raises some interesting questions about how to disentangle the ontological and epistemic features of the bound and related phenomena.
 
Ref \cite{PartI}  largely treated entropy in a classical way,  as a measure of the number of degrees of freedom inside a system or the total amount of information that an observer can get about a system. But of course, our world is really quantum-mechanical in nature and thus ultimately we should be considering quantum von Neumann entropies. This makes the issues discussed in ref \cite{PartI} more complex, because in addition to measuring classical uncertainty, the von Neumann entropy can also measure the entanglement that  a system shares with other systems. Thus in this paper, we set out to  understand how taking entanglement entropy into account affects the interpretation of the covariant bound. In particular, we consider whether interpreting the bound in terms of entanglement entropies may offer novel ways to explain it.

We begin by discussing how the von Neumann entropy fits into the picture, arguing that just as in the classical case, a universal bound on the von Neumann entropy could have either an epistemic or ontological origin. We then note that if the entropies relevant to the covariant bound are entanglement entropies, this opens up new possibilities for explaining the bound as a consequence of features of the entanglement entropy; we discuss several approaches that one might take to do this.  One possibility is to see the bound as a consequence of some kind of entanglement area law; there are some obstacles to be overcome  before this can be achieved, particularly with respect to the interpretation of entanglement entropies of null surfaces, but this may be a fruitful way of proceeding, and in particular may help address some questions left unresolved by ref \cite{PartI} about the relationship between entropy bounds and temporal extension.  We also discuss  the   `spacetime from entanglement'  programme, arguing that entanglement alone may not be able to full ground spacetime topology, but it could potentially play a role in determining the spacetime metric, in which case it would certainly be relevant to explaining the entropy bounds, and this would  potentially support a more ontological reading of the bound.

\section{Background: Entropy Bounds \label{intro}}

Entropy bounds have their origins in an argument due to  Bekenstein\cite{PhysRevD.7.2333} and Hawking\cite{cmp/1103899181} suggesting that if the  second law of thermodynamics is to be obeyed in the vicinity of black holes, we must generalize it to include an entropy associated with black holes equal to $\frac{A}{4   G \hbar}$, where $A$ is the surface area of the black hole. Subsequently discussion on this topic eventually inspired a conjecture known as the `strong entropy bound' which says that the maximum entropy which can be contained in any region of space is upper bounded by $\frac{A}{4   G \hbar}$, where $A$ is the area of any surface bounding the region.  The strong bound  appears to be correct in many physically relevant situations\cite{PhysRevD.23.287}, and it is part of the inspiration for  a flourishing research field aiming to understand quantum gravity in terms of `holography'\cite{Maldacena_1999,https://doi.org/10.48550/arxiv.hep-th/9910146,Kolekar_2010}. 

However, the strong entropy bound is not universally true; it is violated in a number of special cases, some of which, such as inflation and gravitational collapse, are believed to occur in our actual universe.   Thus an alternative known as  the  `covariant entropy bound' (or the `Bousso' bound) has been proposed\cite{Bousso_1999}: instead of bounding the information inside a volume enclosed by some surface, this  version instead bounds the information on a `light-sheet' associated with the relevant surface. A light-sheet is defined as a set of null geodesics leaving the surface orthogonally such that the expansion of the set in the direction going away from the surface is zero or negative, i.e. the geodesics are remaining parallel or coming closer together as we move away from the surface. The light-sheet continues up until the geodesics intersect at a `caustic' (i.e. crossing-point) or encounter a singularity of spacetime. Bousso's bound then simply says that the entropy on the light-sheet associated with a surface of area $A$ is upper bounded by  $\frac{A}{4   G \hbar}$. The covariant bound is currently thought to be   true everywhere in our universe, at least at the semi-classical level.  There also exists a generalized version of the bound\cite{Flanagan_2000} which suggests that the entropy on a partial light sheet bounded on one side by a surface of area $A$ and on the other by a surface of area $A'$ is bounded by $\frac{A - A'}{4   G \hbar}$, but this bound is not universally true, although it has been proven to hold in many physically relevant scenarios\cite{Bousso_2003}.

Moving to the quantum regime, there have been worries that the covariant bound would cease to hold in quantum regimes due to the possibility that quantum fluctuations may violate the null energy condition, meaning that a light-sheet which starts off with negative expansion may briefly experience some positive expansion, causing the information on the light-sheet to become larger than the amount allowed by the covariant bound.  However, ref \cite{Bousso_2014} shows that this will not occur provided that we  require that the non-expansion condition holds everywhere on the light-sheet - then if violations of the null energy condition occur on the light-sheet, one must begin with a strictly negative (rather than zero) expansion, such that even when the expansion increases it still remains less than or equal to zero. This ensures that the von Neumann entropies of the relevant regions will obey the covariant bound   even if the null energy condition is violated (a more general proof which accounts for interactions is given in ref \cite{Bousso_2015}). However,  it remains unclear whether the bound can continue to hold in the regime where quantum gravitational effects become important, for as Smolin notes\cite{Smolin2001TheSA}, in that regime surfaces will not always have a single well-defined light-sheet, so it's unclear how to even formulate the bound.

Ref \cite{PartI} explored the question of how to interpret the covariant bound, distinguishing between `ontological' interpretations of the bound, where it is regarded as an upper bound on the true number of degrees of freedom on  a light-sheet, and `epistemic' interpretations of the  bound, where it is regarded as a constraint on the \emph{accessible} degrees of freedom on a light-sheet. Much of the discussion around the bound and its connection to holography seems to implicitly assume an ontological interpretation of the bound, but there are also good reasons to take the epistemic interpretation seriously. In particular, the  covariant entropy bound is closely related to the `surface postulate,' which is the hypothesis that the number of degrees of freedom available on a surface has an upper bound proportional to the area of the surface. This hypothesis seems independently plausible - certainly, if you have any reason  to believe that the number of degrees of freedom on a surface may be finite, it seems natural to propose that this number may be  proportional to the surface area.

But if the surface postulate is correct, and if information can only be transferred along continuous paths in spacetime, as our best theories of physics seem to suggest, this might be expected to have consequences for the amount of information that observers can extract about the contents of a   region in an infinitesimal interval. In particular, if there are limitations on the rate at which information on the surface can be replaced with new information, then one would expect the surface postulate to induce an upper bound on the information that can be extracted per unit time, which would plausibly be proportional to the area of the bounding surface. And the existence of such rate limitations also seem independently plausible - for example, if one thinks the surface postulate may be a consequence of the discretization of spacetime, then one would presumably expect that discretization to apply to the temporal dimension as well, meaning that information replacement on the surface could not occur infinitely quickly. 

So it would seem quite natural for  the surface postulate to give rise to a bound on the accessible degrees of freedom on a light-sheet, and this bound would have roughly the form of the covariant bound - thus if we accept the surface postulate, an epistemic interpretation of the covariant bound seems quite plausible. Note that this epistemic limitation is epistemic in the precise sense that it describes limitations on information accessible to an external observer, but it may  nonetheless be entirely objective in the sense that those limitations could follow from concrete physical facts about the ontology of the bounding surface.  Ref \cite{PartI}  discussed some arguments for and against both the ontological and epistemic views, and concluded that both possibilities remain open based on current understanding of physics.

 \section{von Neumann Entopies \label{VN}} 

Ref \cite{PartI} largely assumed that the entropies relevant to the covariant bound are classical - in fact, it was argued that these entropies could plausibly be understood either as thermodynamic entropies or informational Shannon entropies, and that either way these entropies could potentially be thought of as encoding   either the total number of degrees of freedom in a region, or the total number of degrees of freedom  accessible to an external observer. 

But the covariant bound is also believed to hold in the context of quantum field theory, and in that context the relevant entropy is the quantum von Neumann entropy\cite{Nielsen}, which is perhaps not so naturally understood as counting degrees of freedom. The von Neumann entropy of a quantum system or spacetime region is defined by $S = -Tr(\rho \ln (\rho))$, where $\rho$ is the density matrix characterising the physical state of the system or region. In this article we will mostly consider cases where $\rho$ is the state of the quantum fields occupying a given region, although in principle one could imagine a generalization in which the state $\rho$  might also include degrees of freedom associated with spacetime itself. It can be shown that the von Neumann entropy  is equal to the Boltzmann entropy and the Shannon entropy in many physically relevant situations, thus justifying the choice to refer to it as an entropy in the first place, but there is some controversy about whether these measures coincide in all possible scenarios\cite{pittphilsci16715,Hemmo2006-MEIVNE,Prunkl2020-PRUOTE}, and  in addition the interpretational questions surrounding the von Neumann entropy are different from those associated with the thermodynamical and Shannon entropies, so we should now consider how the arguments of ref \cite{PartI} might be affected if we stipulate that the entropies of interest are von Neumann entropies rather than classical entropies. 

In fact, up to a certain point, moving from classical to quantum entropies does not significantly change these arguments. This is because, as emphasized in ref \cite{PartI}, the entropy bound is not the value of any specific entropy; rather it is a \emph{universal upper bound} on the entropy, so what matters for understanding its origin is not necessarily the specific nature of the entropies that it bounds, but rather the kinds of physical conditions which could give rise to a universal bound on these entropies.  And  several of the physical conditions which might lead to a  universal upper bound on classical entropies could also be expected to give a  universal bound on the  von Neumann entropy. For example, one of the main arguments for the ontological view of the classical entropy bounds  is the fact that the total number of degrees of freedom in a classical system is an upper bound on its classical entropy, since no physically well-informed observer can assign to a system either a thermodynamical entropy or a Shannon entropy higher than its total number of degrees of freedom. And in the same way, the  dimension of a system's Hilbert space (the quantum analogue of `number of degrees of freedom') is an upper bound on its von Neumann entropy, since no physically well-informed observer can assign to a system a von Neumann entropy greater than the dimension of its Hilbert space. So in the quantum case as in the classical case, one possible explanation for the covariant bound is that it simply counts the total number of degrees of freedom, or equivalently the dimension of the full Hilbert space, in the relevant spacetime region.

Similarly,   ref \cite{PartI} argued that another possible explanation for the classical entropy bound is that it could follow from a universal restriction on the amount of information that external observers can obtain about the content of a spacetime region. This is because  external observers will necessarily describe systems or regions in terms of the number of degrees of freedom that they believe the system or region to have, and therefore if there exist  degrees of freedom that can't be accessed from the outside,  entropy assignations made by external observers will likely be based on a state space which does not include those degrees of freedom, so they will systematically undercount the real degrees of freedom in the region. And much the same argument can be made about the von Neumann entropy: observers will assign quantum states on a Hilbert space defined by the number of degrees of freedom they associate to the system, so again, if there are degrees of freedom that can't be accessed from the outside, their assignations of  von Neumann entropy will likely employ a Hilbert space which does not include those degrees of freedom, and will thus be upper bounded by the size of the accessible Hilbert space rather than the true Hilbert space. 

This shows that the dilemma raised in ref \cite{PartI} carries over to the quantum context: it is not immediately clear whether a universal bound on the von Neumann entropy has an  ontological origin (i.e. it corresponds to the dimension of the actual Hilbert space) or an epistemic origin (i.e. it corresponds to the dimension of the \emph{effective} Hilbert space which is accessible to external observers). 

\subsection{Entanglement \label{entag}}

However, in the quantum case we also have a further issue to consider. For the von Neumann entropy does not just measure classical uncertainty, in the way that classical entropy does: it can also measure the degree of entanglement of a system. That is, the von Neumann entropy is zero for a pure state and non-zero for mixed states, and there are in principle two kinds of mixed states. An improper mixed state is produced when we prepare a system by choosing a pure state according to some probability distribution and then putting the system into that state, so the resulting system is really in some pure state, but someone who doesn't know the outcome of the probabilistic choice cannot tell which pure state in particular it is in. In this case the von Neumann entropy can be regarded as a measure of classical uncertainty about the result of the probabilistic choice, so it is very much like the classical entropy. But a  \emph{proper} mixed state is produced by taking a system which is entangled with some other system and then ignoring (i.e. tracing out) the second system. In this case there are some interpretations of quantum mechanics which would suggest that there is no longer any classical uncertainty, since they hold that the proper mixed state is a complete ontological description of the system, so there is no further fact to be uncertain about. Thus for proper  mixed states the von Neumann entropy may not be so closely analogous to the classical entropy, depending on one's chosen interpretation of quantum mechanics.  Moreover since proper and improper mixed states are represented by the same kind of mathematical object in the density matrix formalism, the way in which von Neumann entropy is calculated makes no distinction between these cases,   so  the entropies relevant to the entropy bound may sometimes measure the degree of entanglement rather than classical uncertainty, or possibly a mixture of both.

Indeed, some physicists take the view that \emph{all} non-zero von-Neumann entropies are a measure of entanglement. For in fact  proper and improper mixed states are closely related: if we assume that unitary quantum mechanics is universal and then write down  a fully quantum description of the unitary operation corresponding to the preparation of an improper mixed state (e.g. a die is rolled and the system is prepared in a pure state depending on the outcome of the roll) we will find that what it actually produces is an entangled state of the die and system, of the form $\sum_i \frac{1}{\sqrt{6}} |D_i \rangle |\phi_i\rangle$ where each $D_i$  corresponds to a possible outcome of the die roll and $\phi_i$ is the corresponding state). Tracing over the die yields a proper mixed state for the system, and this state is mathematically exactly the same as the improper mixed state we would have obtained directly from the probabilistic preparation in the non-unitary description. So if one believes that unitary quantum mechanics is universal, or one subscribes to the church of the larger Hilbert space (which refers to the view that pure state are fundamental and all mixed states are derived from pure states\cite{10.1063/1.1333282}), it follows that all classical uncertainty corresponds to entanglement in this way, so actually \emph{all} mixed states are proper mixed states. Thus our analysis of the entropy bounds should  allow for the possibility that  all of the entropies relevant to the entropy bound might ultimately be entanglement entropies. 

And indeed, a lot of work on entropy bounds in the context of QFT does assume that the relevant entropies are entanglement entropies. For example, this is the approach taken in the quantum proofs of the covariant bound  given in refs \cite{Bousso_2014, Bousso_2015}, in which we define a finite von Neumann entropy on the light-sheet by means of subtracting the vacuum entropy from the matter entropy.  One exception is the alternative quantum proof of the entropy bound due to Strominger and Thompson\cite{Strominger_2004}, further developed in the form of the quantum focusing conjecture by refs \cite{Bousso_2016,ceyhan2019recovering}, which proposes   to address the problem posed by violations of the null energy condition by reformulating the bound  to $S \leq \frac{A}{4 G\hbar} + S'$, where $S$ is the entropy on the light-sheet and $S'$ is the entanglement entropy across the surface. Clearly in this formulation we should not identify the  entropy on the light-sheet $S$ with the entanglement entropy $S'$, as if we do we will end up with the trivially true bound $S' \leq \frac{A}{4 G\hbar} + S'$. Thus the bounds proved in  refs \cite{Bousso_2014, Bousso_2015} and refs \cite{Bousso_2016,ceyhan2019recovering} are inequivalent even in the regimes where they are both well-defined, which is explained by the fact that they define their associated light-sheets differently\cite{Bousso_2016}. It is likely  that  when we move to a strong gravity regime we will ultimately need to use the quantum focusing conjecture rather than the entanglement entropy vacuum-subtraction approach of refs \cite{Bousso_2014, Bousso_2015}, for as noted by Bousso et al\cite{Bousso_2016}, the vacuum subtraction approach only works in a weakly gravitating regime: `\emph{When gravitational backreaction of the state is not negligible, the spacetime geometry is very different from that of a vacuum state. Then it is unclear what one
would mean by restricting both a general state and the vacuum to the ``same” region
or light-sheet. In this case, one cannot define a finite entropy by vacuum subtraction.}' However, since the interpretation in terms of entanglement entropies appears to give the right result in the weakly gravitating regime, we will focus on that interpretation in this article.

And in fact, it turns out that taking the entropies to be entanglement entropies still leads to the same ontic/epistemic dilemma. For on the one hand, the von Neumann entropy is still upper bounded by the dimension of the Hilbert space regardless of whether we understand it in terms of classical uncertainty or entanglement, and therefore a universal bound on \emph{entanglement} entropies could still be ontological in the sense that it might reflect the dimension of the actual Hilbert space.  Yet on the other hand, entanglement can also be associated with knowledge: in particular, if we assume that unitary quantum mechanics is universal and thus  the process of gaining knowledge about a system can be described as a  unitary interaction between the system and the observer, then this interaction will involve the creation of entanglement between the system and the brain of the observer. So under this assumption,  all of the information that an observer outside a region $R$ has about $R$ can be represented as entanglement between the observer's brain and $R$ (although of course the converse is not true - the brain may be entangled with external systems in ways that do not amount to conscious knowledge). Here, we take it that an observer's brain can reasonably be schematized as wholly outside of the region $R$ - this is necessarily an approximation, since the brain itself is a quantum system which will not be perfectly localised, but nonetheless  it appears that  under most circumstances we can think of brains as being fairly sharply localised in well-defined regions. Under this assumption, it follows that the entanglement that the brain shares with the region cannot be larger than the total entanglement of the region with the external world, and thus it is natural to imagine that  the case where the observer has extracted the maximum possible information from that region corresponds to the case in which all of the region's entanglement is with the brain rather than the rest of exterior   - that is, in this case we can think of the brain as simply the compliment to the region. So if there is some epistemic limitation which bounds the total amount of information that observers are able to obtain about the  contents of some spacetime region, then it would be natural for that bound to be manifested in the physics as a bound on the total amount of entanglement that is able to form between this region and its exterior. Thus a universal bound  on \emph{entanglement} entropies could still be epistemic in the sense that it might reflect physical limitations preventing observers from gaining full information about the content of some region. Note that this interpretation of the bound  is not trivial, because in principle one might imagine that the observer could have information which does not take the form of entanglement; but as soon as  we accept that all of the information the observer has can be represented as entanglement between the observer's brain and $R$, it does seem to follow quite directly that a bound on entanglement entropy can be interpreted in an epistemic way.

Indeed, interpreting the entropies in the covariant bound as entanglement entropies may actually strengthen the epistemic account. For a natural concern one might have about the argument presented in ref \cite{PartI} is that the surface postulate only bounds the information passing out of a region in an infinitesimal time interval, so  one might think that   we could eventually gather information exceeding the number appearing in the covariant bound by accumulating information as it emerges over time. This would threaten the epistemic account of the covariant bound: it would entail that if a light-sheet associated with a surface of area $A$ did in fact typically have more degrees of freedom than $\frac{A}{4 \hbar G}$, then we should eventually be able to find out about those degrees of freedom, so  the fact that our current theories  assign  light-sheets associated with surfaces of area $A$ an entropy no more than $\frac{A}{4 \hbar G}$ could not be attributed to  our limited ability to find out about the contents of those regions. 

However, if it is assumed that quantum mechanics is universal and hence information possessed by any observer about the content of the region must ultimately take the form of entanglement between their brains and the region, it follows that the total entanglement entropy between a region and its exterior at a given time  represents an upper bound on the \emph{total} information that has emerged from that region over time. And the purpose of the light-sheet construction is that in ordinary cases a light-sheet will traverse the entire `interior' of a region, as we would normally understand that term. So if it turns out that the entanglement entropy of a light-sheet associated with a surface of area $A$ is typically bounded by $\frac{A}{4 \hbar G}$, then we can argue that in fact the total \emph{accumulated} information available to external observer about the `interior' as captured by a given light-sheet projected from the surface at some time is typically bounded by $\frac{A}{4 \hbar G}$, where $A$ is the surface area at that time. Meanwhile, in cases where the light-sheet fails to traverse the entire `interior' of the region, that is typically because information from the part of the region which does not intersect  the light-sheet cannot reach the surface at all - for example, this often occurs in the presence of singularities, in which case it is known that information content in the surrounding region will remain trapped behind a horizon and thus will not be accessible to external observers. So these results concerning entanglement entropy suggest that the possibility of gathering additional information over time do not in fact undermine the epistemic interpretation of the entropy bounds: there are strict limits on how much information we can obtain about the interior of a region, even if we accumulate information over time.

\section{Explanations}

It appears that interpreting the entropies relevant to the covariant entropy bound as von Neumann entropies does not immediately settle the question of whether the bound is epistemic or ontological in origin. However, it is possible that we can make progress on this question by appealing to the large body of research on features of entanglement entropies in various scenarios. 

At this point we face a difficult problem, which was also   touched on in ref \cite{PartI} - we have available to us many interesting mathematical results establishing connections between entropy, entanglement entropy, gravity, spacetime and so on, but in many of these cases it is not straightforward to determine the correct direction of explanation in a non-question-begging way. One might be tempted to adopt some kind of quietism in which we simply say  there is no fact of the matter about the `correct' direction of explanation. However, this approach may be undermined by the fact that the view we take on these putative explanations seem to suggest different ways of proceeding to formulate theories of quantum gravity and other new physics - for example, in ref \cite{Jacobson_1995} Jacobson argues that gravity may be an emergent phenomenon which is explained by entropic considerations, and that if this is true it would likely have consequences for the correct way to quantize gravity. Thus it is possible in the long term one explanatory approach or another will actually be vindicated by future developments in physics, so  there is a clear sense in which there may actually be a fact of the matter about the correct direction of explanation, and therefore it is worth trying to understand the range of options available to us. 

If the entropies to which the covariant entropy bound applies are indeed entanglement entropies, we have two main options.  First, we could understand the covariant entropy bound as a fundamental fact about ontology or epistemology, and then argue that entanglement entropies are constrained to have certain properties because they have to obey the bound, so these features of the entanglement entropy can be explained by the bound. In this case, it's unlikely that theoretical knowledge about features of the entanglement entropy can tell us much about the nature of the bound, since these features would all  just be consequences of the bound. Or second, we might imagine that   the covariant bound arises as a consequence of certain independent, pre-existing features of entanglement entropies in an underlying quantum field theory or quantum gravity theory, so the bound is explained by the   properties of entanglement entropy, rather than vice versa. In this case, theoretical knowledge about entanglement entropy may indeed help us understand the origin of the bound. 

Thus, although we think both of these options are potentially interesting, for the rest of this article we will mostly focus on the latter approach, seeking to understand what existing theoretical knowledge about entanglement entropies might tell us about the origins of the covariant bound. We will examine several different ways in which one might seek to derive the covariant entropy bound or something similar to it out of features of the entanglement entropy; in each case we aim to understand whether there are good prospects for successfully explaining the covariant entropy bound in this way, and  to determine whether a successful explanation of this kind would support an epistemic or ontological interpretation of the covariant entropy bound.

 \subsection{Area Laws \label{entanglement1}}

 It has been shown that von Neumann entropies in condensed matter systems often obey `area laws'\cite{RevModPhys.82.277,Cramer_2006,PhysRevLett.98.220603,PhysRevLett.94.060503}:  given a discrete condensed matter system which contains mostly short-range interactions (i.e. the Hamiltonian is such that individual particles in the lattice are coupled only with their immediate neighbours, or other close-by particles) we often find that in the ground state most of the entanglement is made up of links between particles at the surface of the system and  particles directly outside the surface, and thus  the entanglement entropy of a subregion scales with the surface area of the region, rather than the volume. 
 A similar phenomenon occurs in quantum field theory\cite{PhysRevD.34.373,Srednicki_1993}: although the full von Neumann entropy suffers from an ultraviolet divergence,   the leading divergence scales as the area of the boundary  between the two regions, because the most entangled degrees of freedom are the high energy ones near the boundary. 

Ref \cite{Chandran_2016} offers a  heuristic illustration of the  origin of such area laws in QFT.  The relevant quantum fields can be decomposed into a set of `modes' of various wavelengths, and  the entanglement entropy counts the modes which cross the boundary, since those are the modes which correspond to entanglement between the two regions.   So roughly speaking, the entanglement entropy is proportional to the number of modes which cross the boundary. Now, if we don't insist on some minimum wavelength for a mode,  then there will be infinite modes crossing the boundary, so this calculation won't give a meaningful result. However, if we impose a minimum wavelength cutoff we will get a finite answer, and as one might expect, the number of modes crossing the boundary is roughly proportional to the area of the boundary. In general  the constant of proportionality in this relationship would be expected to depend on the number of matter fields present, but as shown in ref \cite{Chirco_2014}, since matter fields interact gravitationally, if there are more matter fields we will reach the Planck scale at a higher cutoff length $L$, and in fact   the two effects exactly cancel out, so we end up with a universal relation between area and entropy that does not depend on the number of fields.

Now, clearly area laws are similar in form to entropy bounds, so it is tempting to suppose that the strong or covariant entropy bound may arise as a consequence of area laws. Indeed, it has often been proposed that the Bekinstein-Hawking entropy can be explained by an area law\cite{PhysRevD.77.064013,PhysRevD.34.373,Susskind_1994,jacobson1994black,Frolov_1997} - in fact this was one of Sorkin's main motivations for introducing the concept of `entanglement entropy' in the first place\cite{sorkin20141983} - and since the strong and covariant bounds are essentially generalizations of the Bekinstein-Hawking entropy, it seems reasonable to make a similar proposal here. That said, as argued in ref \cite{PartI}, this approach requires us to accept that the cancellation of the number of fields just happens to work out in the right way, whereas if we take the entropy bound as fundamental we can explain the cancellation as a consequence of the bound in what is arguably a more satisfying and unifying approach, so it should not be taken as a foregone conclusion that the entropy bounds must be explained by area laws and not vice versa. However, for now let us look further into the possibility of explaining entropy bounds in terms of area laws. 

 There are a number of obstacles to be overcome before such an explanation could be given. First, entanglement area laws do not give \emph{universal} bounds on entropy; they simply describe the entropy of some specific state.
 In particular,  area laws are typically derived for the ground state, whereas it can be shown that in the context of many-body systems, a quantum state chosen at random from a Hilbert space will typically obey a volume law rather than an area law\cite{RevModPhys.82.277}. In a continuum quantum field theory on the other hand there is some reason to think that excited states will obey the same area law as the ground state, since it can be proved that at short distances all physical states of a QFT look alike, so they will all have the same structure as the ground state\cite{BUCHHOLZ_1995}. But it is unclear that this result applies generically in the actual world, since  the theory describing the quantum states of spacetime may have a small-distance cutoff, in which case it is not really a continuum QFT - and indeed, as described in ref \cite{Chandran_2016}, the existence of a cutoff is often regarded as part of the explanation for the area law scaling\footnote{Thanks to an anonymous referee for raising this point.}. In addition, even in the ground state, area laws   apply only when the system is not near its quantum critical point\cite{10.1063/1.3554314}. So one might question whether it makes sense to explain the covariant bound, which has been proposed as a \emph{universal} bound on entropy, by appeal to area laws which apply only under quite specific circumstances.   However, perhaps it might be argued that most of the quantum fields we encounter are quite close to their ground state and thus we can expect to find  area-scaling laws applying in a fairly universal way. 

 Second, even in the case where the system in question \emph{is} in the ground state and away from the critical point, the area laws typically apply only to the leading divergence in the calculation of the entanglement entropy, so there can still be subleading terms which could lead to the area bound being exceeded. So if we want to explain the covariant entropy bound in terms of area laws, we would have to consider how significant these subleading terms are, and decide whether small violations of the area scaling are compatible with the existing evidence for the universality of the covariant bound.  

Third, we noted above that in order to obtain a finite value for the entanglement entropy one must choose a UV cutoff, and as ref \cite{sorkin2012expressing} points out, `\emph{When we try to compute $S_{entanglement}$ with finite $\ell$ the answer will depend on the details of how $\ell$ is introduced (sometimes called, rather misleadingly, “scheme dependence”).}' Different ways of introducing the cutoff may may not all give rise to  area laws, or they may give  different constants of proportionality.  So if area laws in QFT are to explain the covariant entropy bound, we would need to show that there is some sensible way of implementing a UV cutoff which typically leads to the right kind of area law, and we would need to find some way of understanding the physical significance of this way of taking the cutoff which explains why the specific scaling behaviour associated with it appears  generic at a macroscopic scale. 

One possibility is to think of the  cutoff scheme as encoding relevant features of the observer, such as the energy scale at which they are able to probe the region. This would align entanglement entropy conceptually with the thermodynamic entropy, whose calculation depends on a choice of coarse-graining which is usually thought of  as encoding relevant features of the observer, such as the set of exogenous variables which they are able to manipulate\cite{maxwell1995maxwell, Myrvold_2020}. In this case, to explain the covariant bound based on area laws it would be necessary to demonstrate that the kind of cutoff scheme associated with typical observers lead to the right kind of area scaling under most circumstances. Alternatively one might think that there is some objectively correct way of taking the cutoff, perhaps corresponding to a real, physical UV cutoff below which QFT ceases to apply\cite{wallace2001defence,binney1992theory,cao1998conceptual,Sakharov:1967pk}. For example,  it has been shown that in the special case of black holes, if we choose the Planck length as the cutoff then calculation of the entanglement entropy across the horizon gives a result proportional to $A$ with a constant of proportionality of the same order of magnitude as the one in the Bekinstein-Hawking formula\cite{sorkin20141983}. In this case, to explain the covariant bound based on area laws it would be necessary to show that the objective physical cutoff leads to the right kind of area scaling under most circumstances. 

What would follow if we could indeed make the case that the  covariant entropy bound is in fact a consequence of entanglement area laws?  First, we emphasize that    the word `law' in the term `area law' is not usually understood as referring to some kind of \emph{fundamental} law; rather `area laws' are   emergent effects arising in certain physical situations as a result of the structure of the underlying QFT, and thus if the covariant entropy bounds is ultimately a consequence of area laws, then it too  should probably be thought of as approximate and emergent rather than fundamental. In particular, we noted in section \ref{entag} that the interpretation of the covariant bounds in terms of entanglement entropies arguably only makes sense in weakly gravitating regimes, so if the bound is explained by area laws then it may only make sense in such regimes - and indeed, Smolin points out that it is in any case unclear how the bound could be properly defined in the strong gravity regime\cite{Smolinholographic}. Second, an account of this kind would clearly favour   an epistemic reading of the covariant bound, since the whole point  of the area laws is  that in many cases the entanglement entropy of a region may be significantly smaller than the real number of degrees of freedom in that region, because the degrees of freedom away from the boundaries don't really contribute.  And hence, since unitary quantum mechanics tells us that  observers external to a region of spacetime  can only gain information about it by being entangled with it,  it follows that observers are limited in their ability to gain information about degrees of freedom deeper in the bulk, leading to an entropy bound which reflects the information accessible to the observer rather than the total information in the region. For example, ref \cite{Hubeny_2007} argues that the entanglement entropy \emph{`provides a measure of how the degrees of freedom localized in that region interact (are ``entangled'') with the rest of the theory. In a sense the entanglement entropy is a measure of the effective operative degrees of freedom, i.e., those that are active participants in the dynamics, in a given region of the background geometry.'}

This kind of `epistemic' account of the covariant bound does not entail that it is in any sense subjective. For facts about the entanglement entropies of ground states are objective physical facts,  and therefore facts about the information accessible to external observers in the relevant kind of systems are also objective physical facts.  Moreover, although it may be possible to interpret cutoffs as pertaining to  capabilities of observers, the `epistemic' account of the bound is prefectly compatible with the idea that there is a real, physical UV cutoff. Indeed, the epistemic account would actually make the relation between a real physical cutoff and the covariant bound more comprehensible - for as noted in ref \cite{PartI}, we might naturally expect that a minimum length cutoff would mean  the information inside a region scales with the volume, but the strong and covariant entropy bounds suggest that it scales with area instead. And the epistemic interpretation in terms of entanglement entropy can  make sense of this fact -  we can now see that even if the total  number of degrees of freedom in a region, as defined by the discretization imposed by the cutoff, does scale with the volume, nonetheless the information \emph{accessible from the outside}, also defined by the cutoff, instead scales with the area, due to the fact that only modes crossing the boundary contribute. So the entanglement entropy calculations provide a useful demonstration of the real and concrete ways in which epistemic limitations on physically embodied observers can come to be reflected in their descriptions of the physical world.

 \subsubsection{Strong vs Covariant Entropy Bounds \label{strong}}

However, accounting for entropy bounds by appeal to entanglement area laws seems most straightforward if we are trying to explain the  \emph{strong} entropy bound rather than the covariant entropy bound. This is because entanglement   is most commonly understood as a relation between two systems \emph{at the same time}, i.e. on the same spacelike surface. So it is quite intuitive to think of the strong bound as bounding entanglement entropies, since it pertains to the content of a region defined on a spacelike slice, and less intuitive to think of the \emph{covariant} bound as bounding entanglement entropies, since it instead pertains to the content of a region defined on a null surface.

 \begin{figure} 
 \begin{center}
\begin{tikzpicture}

\draw (-3.5,-0.5) node  {null}; 
\draw (-2.5,-1.3) node  {spacelike}; 

\draw[gray, thick, ->] (-3,-1) -- (-2.5,0);  
\draw[gray, thick, ->] (-3,-1) -- (-2,-0.7);  
 
\draw[gray, thick] (-1,-1) -- (6,1);  
\draw[gray, thick] (0,2) -- (7,4); 
\draw[gray, ultra thick] (-0.8,-0.5) -- (6.2,1.5); 
\draw[gray, ultra thick] (-0.2,1.5) -- (6.8,3.5)  ; 
\draw (3,2) node  {$B$}; 

 \draw [thick, draw=black, fill=black, opacity=0.2]
       (-0.8,-0.5)-- (6.2,1.5)   --  (6.8,3.5) -- (-0.2,1.5) -- cycle;

\draw (3,1) node  {$B'$}; 
\draw[gray, thick] (-1,-1) -- (0,2);  
\draw[gray, thick] (6,1) -- (7,4);

\end{tikzpicture}
\end{center}
\caption{The light-sheet construction used in  ref \cite{Bousso_2014}}
\label{lightsheet}
\end{figure}
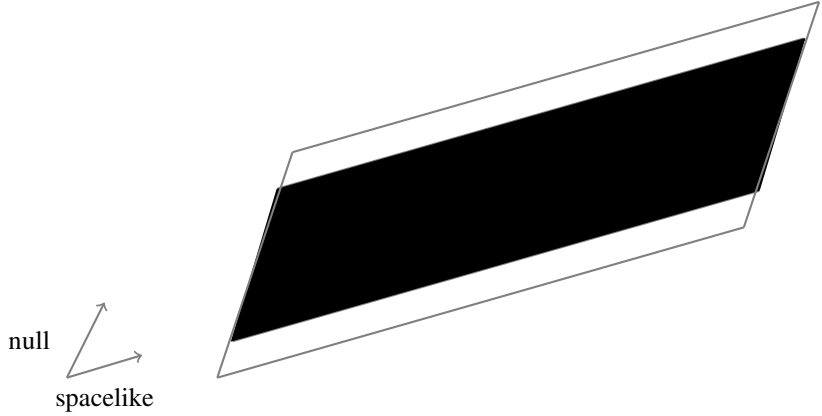

%\filldraw[black] (0,0) circle (2pt) node[anchor=west]{Intersection point};
 
That said, it is possible to   to calculate a von Neumann entropy for a light-sheet, as done in ref \cite{Bousso_2014}, which aims to prove a quantum version of the generalized entropy bound as discussed in section \ref{intro}. The calculation proceeds by extending the light-sheet associated with the surface $B$ to a larger null surface, as depicted in figure \ref{lightsheet}, then defining a vacuum state and the actual state of matter on this null surface, then restricting them to the part of the null surface occupied by the light-sheet portion bounded by surfaces $B$ and $B'$, thus obtaining reduced density matrices for the vacuum and actual matter state on the light-sheet. The von Neumann entropies of these density matrices diverge, but the difference between them is finite, and it is this value which is shown to be proportional to $\frac{A}{4 \hbar G}$. (A more complex calculation allowing for interacting fields is performed in ref \cite{Bousso_2015}).  

However, can the result of this calculation can   be interpreted as encoding the amount of entanglement between the light-sheet and its exterior region? First of all, it's not necessarily obvious which exterior region we are talking about. In the ordinary case when we define entanglement entropies we are looking at a region defined on a spacelike slice entangled with another region on the same spacelike slice, so there is no question about the time. But the light-sheet is not defined on a single spacelike slice, so we can't simply take that route. We could perhaps imagine integrating over a stack of spacelike slices which together foliate the whole light-sheet, but it's unclear this would make sense - the resulting calculation of the entropy for the exterior region  would surely end up vastly overcounting the entropy, since the same entropy-bearing systems would be integrated over many times. But it would seem arbitrary to select just \emph{one} spacelike slice. So in fact, probably the most the natural approach is to extend the null surface on which the light-sheet is defined beyond the bounding surfaces $B, B'$ and then consider the light-sheet to be entangled with the region outside of the surfaces \emph{on this null surface} - and indeed, this appears to be the interpretation suggested by the calculation that is actually performed in ref \cite{Bousso_2014}, since as we have noted, it starts by defining a state for an extended null surface and then restricting it to the light-sheet. 

So, can we just say the von Neumann entropy of the light-sheet quantifies its degree of entanglement with the rest of the null surface? This isn't so obvious, since the rest of the null surface is in the (lightlike)  \emph{future} and/or \emph{past} of the light-sheet, so we would not normally say that  it is related to the light-sheet \emph{only} by entanglement. Of course, there are certainly ways in which these regions could be connected by what we would normally think of as entanglement.   For example, consider the case where $S$ is a sphere and the backwards light-sheet of $S$ traverses the entire interior region of the sphere. Although of course this is a relativistic scenario which should properly be associated with a relativistic quantum field theory, let us briefly switch into non-relativistic quantum mechanics, which offers a more intuitively accessible language for describing entanglement phenomena.  So imagine   that there is an entangled pair of particles present, with one particle on the inside of the sphere and the other particle on the outside. If neither particle interacts with anything during the relevant time interval, then there is a sense in which we could say that the interior particle is, at the time when the light-sheet intersects it, entangled with the exterior particle not only on the same spacelike slice but also on all future spacelike slices, until the exterior particle undergoes some interaction which  breaks the entanglement or transfers it elsewhere. After all, the predictions of quantum mechanics for the outcomes of measurements on these particles are the same regardless of whether we measure them on the same spacelike slice or the same null surface, so it makes sense to think of the latter case as instantiating a kind of temporal entanglement, which is really just the temporal development of pre-existing spatial entanglement. Thus in this case the interior particle is `entangled' with the future version of the exterior particle which intersects with the relevant null surface extended outside of the sphere. And, if we now switch back into the domain of relativistic quantum field theory,   it seems reasonable to think that  something similar could be said about the quantum fields in these regions, allowing us to say that two regions on the same null surface can indeed be entangled with each other. 

However, unlike in the spacelike case,  entanglement is not the only kind of direct connection we can have between these two regions, because the lightlike relation between the regions allows for physical systems to cross the boundary carrying information directly from one to another - in particular, if we are considering a past-directed light-sheet, so the  exterior region outside of the surface B lies in the future of the light-sheet, we can have physical signals crossing the boundary to carry information  out of the light-sheet into the exterior, and if we are considering a future-directed light-sheet, so the  exterior region outside of the surface B lies in the past of the light-sheet, we can have physical signals crossing the boundary to carry information from the exterior into the lightsheet.  So the correlations between the regions may be due to pre-existing entanglement, but may also be due to physical signals passing in and out of the light-sheet. 

How does this second kind of connection relate to the von Neumann entropy on the light-sheet calculated in ref \cite{Bousso_2014}? Well, the interpretation of this entropy is not very clear, but it seems reasonable to think that the calculated entropy may  include both kinds of correlations  -  i.e. in some sense it quantifies both connections associated with pre-existing entanglement and also connections associated with physical signals traveling through the boundary\footnote{Perhaps  one could make sense of what is going on here in the context of an approach akin to the proposal of Aharonov, Popescu and Tollaksen\cite{AharonovPopescuTollaksen}, who argue that time evolution can be thought of as a kind of `entanglement' correlation between Hilbert spaces representing successive moments of time. This would be an interesting way to unify `pre-existing entanglement' with `physical signals traveling through the boundary' but it remains to be seen if it would give the right quantitative results.}.  After all, physical signals crossing the boundary are necessary to \emph{produce} entanglement in the first place, because entanglement cannot be created at a distance\cite{PhysRevA.53.2046}. That is, given an initial state in which a pair of systems are not originally entangled, in order for them to later become entangled, we must either perform a joint quantum operation on them before separating them, or we must entangle a different pair of systems  by a joint quantum operation and then transfer the entanglement from one pair of systems to another by a process like entanglement swapping\cite{PhysRevLett.71.4287}. Therefore any entanglement between the inside and outside of a region must either have existed since the relevant initial condition - i.e. presumably since the beginning of time -  or it must have originated from some physical system which passed through the boundary at some point. And since we have no obvious way to determine whether or not entanglement in the present has its origins in an initial condition or was created in a later interaction, it's unclear that one could ever really separate out entropy coming from entanglement and entropy coming from physical systems crossing the boundary - and if we were to try, it seems quite likely that one would find the distinction to be reference-frame dependent. So it seems reasonable to think that the von Neumann entropy calculated in ref \cite{Bousso_2014} includes or perhaps even is dominated by contributions  from physical signals crossing the boundary.

This means that  the calculation of ref \cite{Bousso_2014} could be interpreted as offering a quantum version of the epistemic account of the covariant bound previously suggested in ref \cite{PartI}, i.e. the idea that the  bound originates from an epistemic restriction due to the fact that the amount of information which can pass through a bounding surface in an infinitesimal interval is upper bounded by the number of degrees of freedom  on the surface. Thus in the case of the covariant bound, it doesn't seem to be the case that invoking area laws would \emph{replace} or make redundant the kind of epistemic explanations we have previously considered - the two approaches are quite complementary and  the entropy calculation on a null surface may potentially be regarded as a more formal expression of the epistemic restrictions motivated in an intuitive way in ref \cite{PartI}.

Of course,  we do still have area laws for areas defined entirely on spacelike surfaces which presumably include only contributions from entanglement, without any physical information flow from one region to the other, and one might worry that if these pure entanglement area laws are not the explanation of the covariant bound, then it is then a mysterious coincidence that  the entropy featuring in the covariant bound and the entropy featuring in spacelike area laws both scale with the surface area. But in fact there is no coincidence here. The entanglement in the area laws is related to the surface area because entanglement is created by local interactions through the interfacing surface, and the entropy associated with a light-sheet is related to the surface area because both the creation of entanglement and information passing out of the region must be mediated by local interactions through the interfacing surface. So, as suggested by ref \cite{RevModPhys.82.277}, rather than a coincidence we may have something like a common cause: both phenomena scale with the surface area due to the locality of interactions through an interfacing surface.

 \subsection{Spacetime \emph{From} Entanglement \label{from}}

 A more radical approach to connecting entanglement entropy area laws with entropy bounds is based on the conjecture that the structure of spacetime is in some sense derived from entanglement.  It has been suggested that rather than explaining area laws in terms of the locality of interactions in spacetime, we should instead derive the structure of spacetime from entanglement\cite{Maldacena_2013,VanRaamsdonk:2010pw}:  for example ref \cite{Cao_2017} conjectures that `\emph{mutual information ... can be used to associate spatial manifolds with certain kinds of quantum states}' (in this context the mutual information can be used as a measure of entanglement entropy). The idea is that the facts about where an object is located in spacetime are to be derived as a function of its degree of entanglement: `\emph{degrees of freedom are only “in the interior” in a geometric sense when they are entangled with their neighbors but not with distant regions.}' It seems plausible that this approach could offer an   explanation of the covariant bound, since clearly if spacetime itself is derived from entanglement there will necessarily be strong links between spacetime structure and von Neumann entropy.

Now, the `spacetime from entanglement' approach has been criticized in a number of ways - for example, as   noted by Ney\cite{NeyForthcoming-NEYFQE},  if the entangled objects are not located in a pre-defined spacetime, it's  unclear why the entanglement they exhibit should have exactly the structure which is compatible with our ordinary three-dimensional space:  to give rise to a 3D space the state must belong to the space of `redundancy-constrained' states, and ref \cite{Cao_2017} emphasizes that these states are not at all generic in the space of possible states, so it seems we must simply impose this structure  in an ad hoc way\footnote{Though of course, proponents of this view might respond that simply imposing a 3D structure on spacetime itself is equally ad hoc!}. But here we want to focus on a different kind of objection. First, it is important to be clear about exactly what is meant by the claim that `spacetime' can be derived from entanglement. Spacetime has both topological features and metrical features, where the \emph{topology} of spacetime refers to properties that are invariant under any continuous deformation of the spacetime, such as facts about which points are adjacent, while the metric provides a notion of distance between the points in the space, and thus pertains to properties that are not invariant under continuous deformations of the spactime. It seems that ref \cite{Cao_2017} ultimately aims to extract both the topology and the metric of spacetime from facts about entanglement - after all, the suggestion that `\emph{degrees of freedom are only “in the interior” in a geometric sense when they are entangled with their neighbors but not with distant regions,}' seems to make sense only if we accept that not only the metric but also topology is determined by entanglement, since   whether or not a point is inside of a region is a topological fact. Thus an appropriate way to assess the success of this approach is to consider  whether or not the structure thus derived is capable of performing all the characteristic functions of both spacetime topology and the spacetime metric.

In particular,   one key functional role of the topology of spacetime is  that it determines which systems can interact -  where by `interact' we mean specifically  processes involving the exchange of information, i.e. the kind of processes that could be used for signalling. According to    our best current understanding of physics, such processes always occur locally in spacetime, meaning that information transfer can only occur between systems which are adjacent: this feature of physics is sometimes known as `no-signalling,' or `locality' or `causality' and it features prominently in quantum mechanics, QFT and special relativity. Moreover, `causality' is not just an arbitrary feature that these theories happen to have: for if superluminal signalling were possible then in principle we would be able to use it to construct closed causal loops and causal paradoxes and thus to create logical inconsistencies\cite{QMG}, so there are reasons that go beyond  any of these individual theories to think that the locality of interactions must be a deep and significant feature of our reality. And to enforce local interactions one needs at least adjacency relations encoded in a topological manifold, and possibly also one also needs conformal structure, depending on whether one expects the lightcone structure of spacetime to be determined as part of the locality constraint or whether it is expected to emerge in some more dynamical way. Thus the role played by the topology of spacetime in defining adjacency relations, and/or the role of conformal structure in defining possible signalling relations, is one of its most important functions. Moreover, we have already noted that covariant bound appears to be closely related to the locality of interactions across surfaces, so the role of spacetime topology and/or conformal structure in defining adjacency relations is particularly relevant if we are seeking to explain the bound. 

However, it seems unlikely that spacetime could play this role if its topology and/or conformal structure were in fact derived entirely from entanglement and nothing else.  It is of course possible to get adjacency relations from the approach of ref  \cite{Cao_2017} -  specifically,  the authors first use the proposed relation between mutual information and distance to assign distances to pairs of regions, and then use multidimensional scaling to embed them in a manifold, and thus, because they have arrived at a manifold at the end of this process, they have necessarily also arrived at  a notion of adjacency. However, it is a little unclear how adjacency could come to have the dynamical significance with respect to interactions that it in fact has if it were solely grounded on entanglement: for in this picture `adjacency' essentially just corresponds to being highly entangled, but being highly entangled doesn't usually have anything to do with being able to exchange information. Indeed, the `no-signalling' theorem in quantum mechanics refers precisely to  the fact that information \emph{cannot} be exchanged using entanglement alone, so it would be surprising if being highly entangled suddenly made systems capable of exchanging information\footnote{Recall that the `no-signalling theorem' simply refers to the fact that measurements performed on distinct quantum systems commute, and therefore classical information cannot be sent from one system to another using entanglement, regardless of when and where in spacetime they are located. Thus this use of the no-signalling theorem does not presuppose the existence of conformal structure - it still makes sense in a pure topological manifold.}. Furthermore, `degree of entanglement' is a continuous measure, whereas `adjacency' is not - either two systems are adjacent and can interact, or they are not adjacent and cannot interact - so although the procedure of ref   \cite{Cao_2017} does arrive at a notion of adjacency by creating an embedding based on entanglement, from a dynamical point of view it seems hard to see how a sharp distinction which is not present in the original entanglement structure but which is merely a byproduct of the mathematical procedure of creating an embedding could suddenly come to have so much dynamical significance in determining how the system can interact. And finally,  there seems to be a potentially problematic circularity in this vicinity, because  one of the distinctive features of entanglement is that local interactions are required to \emph{create} entanglement, and it seems difficult to make sense of this constraint if we derive spacetime structure entirely from entanglement: surely there needs to be some pre-existing spacetime structure to determine how that entanglement can come into being in the first place? This point is related to Ney's worry in ref \cite{NeyForthcoming-NEYFQE}, for the usual  way of explaining why entanglement in our world has the structure that it does is to suppose that it is produced in  a dynamical process involving local interactions in some pre-existing spacetime, but we will not be able to explain it this way if we take it that entanglement is ontologically prior to facts about locality in spacetime, which leads to a version of  the dilemma posed by Ney:  if spacetime structure emerges from entanglement and not vice versa, it seems mysterious that the entanglement in our world should have the specific kind of structure that \emph{would} have been produced by dynamically local interactions in a four-dimensional spacetime, even though it cannot in fact have been produced that way.

So in our view, it is unlikely that \emph{all} important facts about spacetime can be derived entirely from entanglement. That said, there may be a viable intermediate position: although it seems  unlikely that entanglement alone can fully  define the topology of spacetime,  entanglement could potentially be used to define the \emph{metric} of spacetime. That is, rather than first obtaining distance relations from entanglement and then using these to obtain an embedding on a manifold as suggested in ref \cite{Cao_2017}, we could instead \emph{start} from a bare topological manifold or conformal manifold with no metric, with quantum systems living at each  point of the manifold; and then,   using the facts about adjacency defined by the underlying manifold, allow entanglement to form by means of local interactions between adjacent systems. (Or alternatively, we could start from kind of non-spatiotemporal substratum which is not to be identified with spacetime but which does define some standard of adjacency - for example the edges in loop quantum gravity\cite{cc}). And we could then  postulate that the \emph{metric} of the spacetime is subsequently defined by facts about entanglement, for example by specifying that the distance between two regions (a metrical feature) is given by a function of the quantum mutual information between the regions. 

The idea that entanglement may  define only the metric and not the topology of spacetime is not necessarily a problem for the `spacetime from entanglement' program, for it would still be very interesting if we could understand the spacetime metric as a function of entanglement relations. It would however presumably entail that the covariant bound cannot be defined entirely by reference to entanglement entropies, for the bound clearly has something to do with topology - the whole point of the light-sheet construction is to give a covariant way of identifying the \emph{interior} of a region, and the notion of an interior is at least in part a topological notion. But nonetheless, if entanglement did play a role in defining the metric of a spacetime, then it would certainly have a close connection to the covariant entropy bound, since area is determined by the metric and hence it is the metric which gives us the area $A$ appearing in the  bound. For example, if one were to say that the area of a surface is, by definition, equal to some constant times the entanglement entropy passing through it, then it would be true by definition that the entanglement entropy of some region must have a bound proportional to its surface area. 

What would such an account mean for the epistemic/ontic distinction? On the one hand, it is still true in this picture that the entanglement entropy of a region  quantifies the total amount of information that external observers can have about that region; so if we then understand the covariant bound as a consequence of a bound on the  entanglement entropy, it would not be wrong to say that the covariant bound reflects some kind of epistemic limitation.  However, in the spacetime-from-entanglement approach the entanglement entropy is  also understood as  defining the ontology of spacetime, so it's not clear that it still makes sense in this picture to say that there  there can be more degrees of freedom inside a region than external observers are able to access - in a sense, the spacetime-from-entanglement approach blurs or erases the distinction between epistemic and ontic accounts of the ontology of spacetime. Thus the approach might be most compatible with some kind of relational or perspectival picture in which we say that the  `number of degrees of freedom' of a region can only be defined relative to an external observer, in which case the maximum  entanglement entropy of a region is by definition equal to the total number of degrees of freedom in that region. 

Moreover, as noted in ref \cite{PartI}, one might  think that the ontology of spacetime \emph{should} be defined in such a way that the total information content closely matches the accessible information. After all, our intuitive picture of spacetime as a continuum completely filled up with degrees of freedom has the consequence that the ontology of spacetime far outstrips the evidence available to us, which seems problematic if we value ontological parsimony, and in combination with the surface postulate apparently  leads to the troubling possibility   that there could exist whole  realms of physics which are permanently inaccessible to us.   So if it can be shown that a `spacetime-from-entanglement' approach does indeed help close the gap between the `true' information content of spacetime and the accessible information, that might be an argument in its favour.

Before closing our discussion of `spacetime from entanglement' let us mention one obvious objection to the conjecture that the metric of spacetime arises out of entanglement structure. The argument of ref \cite{Cao_2017} show only how to use entanglement structure to define spatial distances, not temporal intervals; and yet  in Special and General Relativity the spacetime metric defines  both spatial and temporal distances. The authors of ref \cite{Cao_2017} suggest that their approach to defining the geometry of space at a single time could be combined with time evolution to describe the geometry of spacetime as a whole, but this seems somewhat in tension with the underlying principles of special relativity: for in a relativistic context  local Lorentz transformations can be used to transform a spatial interval into a combination of spatial and temporal intervals, so one might naturally wonder how deriving spacetime from entanglement could possibly be compatible with relativity if this approach requires us to treat spatial and temporal distance as two entirely different categories with different underlying ontologies. Of course, enacting a split between space and time is  compatible with a Hamiltonian formulation of general relativity, so if there were a natural way to derive the constraint equations of the Hamiltonian formulation this approach might be made at least formally compatible with relativity\footnote{Thanks to an anonymous referee for raising this point.} Still, since   this approach would presumably require us to choose some privileged foliation of spacetime, it would arguably still be somewhat unappealing from a relativistic standpoint. In addition, note that every entangled state is a quantum superposition of unentangled states, so the linearity of quantum mechanics would suggest that the dynamics of an entangled state should simply be a linear combination of the dynamics for the unentangled states. Thus if entanglement in and of itself is capable of producing dynamical effects, this would potentially violate the unitary of quantum mechanics, since dynamics due specifically entanglement would presumably not be expressible as a linear combination of unentangled dynamics\footnote{Thanks to an anonymous referee for raising this point.}.

However, it is possible that rather than defining spatial geometry at a point and then performing time-evolution, one could use methodology similar to ref \cite{Bousso_2014} to define entropies on null surfaces, or methodology similar to refs \cite{sorkin2012expressing,Chandrasekaran_2023} to define entropies in 4D regions, thus giving something like `entanglement entropy' which could be used to define timelike and lightlike distances as well as spacelike ones. But as noted in section \ref{strong}, it's unclear that these kinds of entropies measure only `entanglement' in the usual sense of the word - they would seem to include also physical signals passing between regions, so this might require us to change the interpretation of the construction in some way. Thus as in section \ref{strong}, it seems that improving our understanding of the correct interpretation of von Neumann entropies defined for null surfaces or 4D regions would be useful to help elucidate the connection between entanglement and entropy bounds.

 \section{Conclusion} 

 In this article we set out to understand how to interpret the covariant entropy bound in a context where the entropies bounded may be be  quantum von Neumann entropies rather than classical entropies. Our initial observation is that both the ontological and epistemic interpretations of the bound remain possible, and we also noted that this conclusion holds regardless of whether the von Neumann entropy is understood as encoding classical uncertainty, entanglement entropy, or any combination of the two. 

Now, if the entropies relevant to the covariant bound are   entanglement entropies, one possible option would be to regard the covariant bound as a fundamental fact about ontology or epistemology, which would allow us to use the bound to explain various features of entanglement entropies in the underlying quantum field theory or gravitational theory. But alternatively, one might also suppose that the covariant bound should be viewed as emerging from deeper features of entanglement entropy;  in this article we have discussed several ways in which one might try to make that case.  First we saw that entanglement area  laws could potentially offer what appears to be an epistemic explanation of the covariant bound, although this depends on some open questions about how to interpret entropies defined on null surfaces. An explanation by appeal to properties of the entanglement entropy might end up being closely related to the intuitive `epistemic'  approach suggested in ref \cite{PartI}, since both seem closely tied to facts about the locality of interactions through boundaries. 
 
 Next, we discussed the idea that spacetime structure is derived from entanglement. We argued that although entanglement might potentially be the origin of the metric of spacetime, it is less plausible that it   is  also the origin of the topology. But nonetheless, we concluded that the spacetime-from-entanglement approach might point to a more ontological way of thinking about the bound, since in this picture   entanglement entropy defines the ontological content of spacetime.

Overall, our conclusion is that even if the covariant bound is ultimately a consequence of properties of the entanglement entropy, both ontological and epistemic interpretations of it remain possible. Current established knowledge about entanglement entropies  seems to point  towards an epistemic interpretation of the  bound, but this could change if the `spacetime from entanglement' approach were to gain more mainstream acceptance. In any case, if theoretical knowledge about entanglement entropies is indeed pointing us towards an epistemic interpretation of the bounds, that seems interesting and significant, due to the close connection between the covariant bound and gravity; the role of entanglement entropies here may perhaps point to a novel way of understanding gravity in the context of quantum mechanics. 

\section{Acknowledgements}

Thanks to Yasaman Yazdi for very helpful comments on a draft of this paper.

 \end{document}